\documentclass[twocolumn,prl]{revtex4}
\usepackage{latexsym}
\usepackage{dcolumn}
\usepackage{bm}

\newcommand{\lista}[2]{\newcounter{#1}\begin{list}
{$\bf #2_{\arabic{#1}}$}{\usecounter{#1}}}

\begin{document}
\title{Note on Possibility of obtaining a non-relativistic proof of the
spin-statistics theorem in the Galilean frame}

\author{Gabriel D. Puccini}
\email{gpuccini@umh.es} \affiliation{Instituto de Neurociencias,
Universidad Miguel Hern\'andez, Apartado 18, 03550 San Juan de
Alicante, Spain.}

\author{H\'ector Vucetich}
\affiliation{Observatorio Astron\'omico, Universidad Nacional de
La Plata, Paseo del Bosque S./N., (1900) La Plata, Argentina}

\begin{abstract}
We replay to the critique by Sudarshan and Shaji of our argument
of impossibility to obtain a non-relativistic proof of the
spin-statistics theorem in the Galilean frame.
\end{abstract}

\maketitle

In a recent note \cite{Suda04}, Sudarshan and Shaji have presented
one objection to the argument of impossibility to obtain a
non-relativistic proof of the spin-statistics theorem suggested by
us in \cite{Pu04}. To solve the incompatibility between Hermitian
field operators and Galilean invariance for massive fields,
Sudarshan and Shaji assert that hermiticity can be accomplished by
doubling the number of components of $\hat \xi_\lambda$ and
choosing $M$ as:
$$
M = m \pmatrix{0&-i\cr i& 0} \;\;.
$$

Finally the authors conclude that ``The essential point is that a
field (of any spin) can be made to carry an additional charge by
doubling the components while still keeping them real. The `mass'
$M$, which may be considered as just another charge, can also be
accommodated in an identical fashion. To lament over the Bargmann
phase due to $M$ and not worry about any other charge in relation
to the proof of the spin-statistics connection stems from
assigning $M$ a special status over any other charge that may be
relevant to the fields that are being considered''.

Doubling the number of components of the field operators, as is
proposed by Sudarshan and Shaji, is the natural way of introducing
the concept of charge in the Schwinger formalism for relativistic
quantum field theory \cite{Sc51,Sc53}. To see an example coming
from this theory, we suppose that we double the number of
components of a given field operator
and we choose the charge matrix representation as:

$$
Q = e \pmatrix{0&-i\cr i& 0} \;\;.
$$

The resulting hermitian field operator will then describe a
charged field composed of particles with charges $+e$ and $-e$
(the eigenvalues of $Q$) \footnote{The evident symmetry between
particles and antiparticles can be shown more clearly by writing
$Q$ in diagonal form, but the hermiticity of the fields is lost.
This means that in the relativistic theory, the possibility of
working with hermitian or non-hermitian field operators can be
decided by choosing an appropriated base.}. But the price for a
such simple introduction of charge is the implicit assumption of
crossing symmetry between particles and antiparticles.

It is in this point where relativistic and galilean quantum field
theories show their most important differences. Indeed, mass $M$
is a charge in galilean field theories, but it cannot be treated
in the same way as in the relativistic theory because crossing
symmetry is not required. In other words, particles ($+m$) and
antiparticles ($-m$) cannot be considered on the same footing.
This can be easily illustrated for the case of spin-zero field
\cite{JMLL67}. If we assume the usual commutation or
anticommutation rules for the annihilation and creation operators
of particles and antiparticles,
$$
[\hat a(k'),\hat a^{\dag}(k)]_{\mp}= \delta (k'-k) \;\;\;,\;\;\;
[\hat b(k'),\hat b^{\dag}(k)]_{\mp}= \delta (k'-k)
$$
and we construct a field operator with the correct galilean
transformation properties by taking linear combinations of
particles annihilation operator and antiparticles creation
operator as:
\begin{eqnarray*}
\hat \chi({\bf x},t) = (2\pi)^{-3/2} \int d\mu(k) [ \alpha
e^{\frac{i}{\hbar}(Et-{\bf p.x})} \hat a(k) +\\
\beta e^{-\frac{i}{\hbar}(Et-{\bf p.x})} \hat b^{\dag}(k)]\;,
\end{eqnarray*}
we arrive to the following commutation or anticommutation rule:
$$
[\hat \chi({\bf x},t) ,\hat \chi^{\dag}({\bf y},t)]_{\mp}=
(|\alpha|^2 \mp |\beta|^2 ) \delta^3 ({\bf x}-{\bf y})\;\;,
$$
which is satisfied for any value of $\alpha$ and $\beta$. In
particular equal contribution of particles and antiparticles with
$\alpha=\beta$ implies that the commutator vanishes identically.
It should be note, moreover, that no spin-statistics relation can
be deduced. A complete proof of these two important results of the
galilean theory for any spin can be found in \cite{Pu01}.

In conclusion, it is not possible to double the number of
components of the field operators because it implies to assume
equal contribution of particles and antiparticles. And, as we saw,
this powerful result of the relativistic theory is not more valid
in the galilean theory.

\end{document}